\documentclass[acmsmall,nonacm]{acmart}
\usepackage{subfig} 
\usepackage{enumerate}
\usepackage{paralist}
\usepackage{multirow}
\usepackage{amsmath}
\usepackage{url}

\AtBeginDocument{%
  \providecommand\BibTeX{{%
    \normalfont B\kern-0.5em{\scshape i\kern-0.25em b}\kern-0.8em\TeX}}}

\makeatletter
\let\@authorsaddresses\@empty

\newcommand{\fakepar}[1]{\vspace{0mm}\noindent\textbf{#1.}}
\newcommand\figref[1]{Fig.\,\ref{#1}}
\newcommand\secref[1]{Sec.\,\ref{#1}}
\newcommand\tabref[1]{Tab.\,\ref{#1}}
\renewcommand{\eqref}[1]{Eq.\,\ref{#1}}

\usepackage{acronym}
\acrodef{STT-MRAM}[STT-MRAM]{Spin-Transfer Torque Magnetic Random-Access Memory}
\acrodef{NVM}[NVM]{Non-Volatile Memory}
\acrodef{AC}[AC]{Approximate Computing}
\acrodef{IC}[IC]{Intermittent Computing}
\acrodef{QoR}[QoR]{Quality of Results}
\acrodef{ACT}[ACT]{Approximate Computing technique}
\acrodef{ARE}[ARE]{Average Relative Error}
\acrodef{RMSE}[RMSE]{Root Mean Squared Error}
\acrodef{PSNR}[PSNR]{Peak Signal-to-Noise-Ratio}
\acrodef{CRC}[CRC]{Cyclic Redundancy Check}
\acrodef{NTC}[NTC]{Near Threshold Computing}
\acrodef{MCU}[MCU]{microcontroller unit}
\acrodef{NN}[DNN]{Deep Neural Network}
\acrodef{IOT}[IoT]{Internet of Things}
\acrodef{WER}[WER]{Write Error Rate}
\acrodef{SRAM}[SRAM]{Static Random Access Memory}
\acrodef{QL}[QL]{Quality Level}
\acrodef{FRAM}[FRAM]{Ferroelectric Random Access Memory}
\acrodef{DMA}[DMA]{Direct Memory Access}

\begin{document}

\title{Energy versus Output Quality of Non-volatile Writes in Intermittent Computing}

\author{Rei Barjami}
\affiliation{%
	\institution{Politecnico di Milano}
	\country{Italy}
}
\author{Antonio Miele}
\affiliation{%
	\institution{Politecnico di Milano}
	\country{Italy}
}
\author{Luca Mottola}
\affiliation{%
	\institution{Politecnico di Milano and RI.SE}
	\country{Italy, Sweden}
}

\renewcommand{\shortauthors}{Barjami et al.}

\begin{abstract}

We explore how to improve the energy performance of battery-less \ac{IOT} devices at the cost of a reduction in the quality of the output.
Battery-less \ac{IOT} devices are extremely resource-constrained energy-harvesting devices.
Due to erratic energy patterns from the ambient, their executions become \emph{intermittent}; periods of active computation are interleaved by periods of recharging small energy buffers.
To cross periods of energy unavailability, a device persists application and system state onto \ac{NVM} in anticipation of energy failures.
We purposely control the energy invested in these operations, representing a major energy overhead, when using \ac{STT-MRAM} as \ac{NVM}.
As a result, we abate the corresponding overhead, yet introduce write errors. 
Based on $1.9+$ trillion experimental data points, we illustrate whether this is a gamble worth taking, when, and where.
We measure the energy consumption and quality of output obtained from the execution of nine diverse benchmarks on top of seven different platforms.
Our results allow us to draw \emph{three key observations}: 
\begin{inparaenum}[i)]
\item the trade-off between energy saving and reduction of output quality is \emph{program-specific}; 
\item the same trade-off is a function of a \emph{platform's specific compute efficiency and power figures}; and 
\item \emph{data encoding and input size} impact a program's resilience to errors.
\end{inparaenum}
As a paradigmatic example, we reveal cases where we achieve up to 50\% reduction in energy consumption with negligible effects on output quality, as opposed to settings where a minimal energy gain causes drastic drops in output quality. 
\end{abstract}


\maketitle

\acresetall{}

\section{Introduction}

Ambient energy harvesting allows \ac{IOT} devices to eliminate their dependency on traditional batteries~\cite{harvesting-survey}.
This enables drastic reductions of maintenance costs and previously unattainable deployments, operating completely unattended for multiple years~\cite{water-deployment-microbial-fuel-cell,tethys,soil-termoelectric,sensys20deployment,denby2023kodan}
However, energy from the environment is generally erratic, causing frequent and unanticipated energy failures.
For example, harvesting energy from RF transmissions to compute a simple CRC application may lead to $16$ energy failures over a $6$ second period~\cite{harvesting-survey}.
Executions thus become \textit{intermittent}, as they consist of intervals of active operation interleaved by possibly long periods of recharging energy buffers~\cite{harvesting-survey}.

\begin{figure}[t] 
    \includegraphics[width=0.60\linewidth]{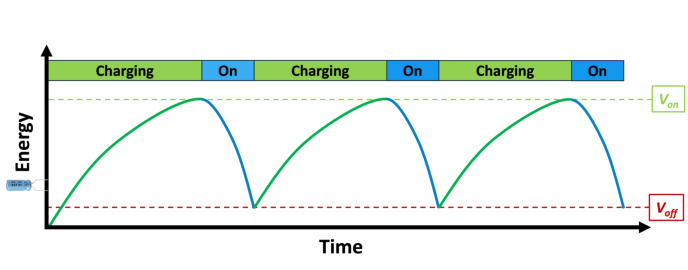}
    \caption{Intermittent execution.}
    \label{fig:intermittent}
\end{figure}

\fakepar{Computing intermittently} \figref{fig:intermittent} shows an example execution.
The ambient charges an onboard small energy buffer, such as a capacitor, until the energy buffer voltage level~$V_{\mathit{on}}$ is reached, causing the device to power on.
The device senses, computes, and communicates as long as the energy buffer voltage level charge remains above a threshold $V_{\mathit{off}}$.
The device then switches off, waiting for the energy buffer to reach $V_{\mathit{on}}$ again.

Battery-less \ac{IOT} devices are \emph{extremely resource constrained}~\cite{batteryless-future}: they typically feature 16- or 32-bit \acp{MCU} with kilobytes of main memory.
Applications run on bare hardware \emph{without operating system support}.
Because of this, energy failures normally cause a device to lose application and system state.
To ensure forward progress across energy failures, a variety of techniques exists that make use of \emph{persistent state} stored on \ac{NVM}, as we elaborate in~\secref{sec:related}.
Persistent state is loaded back from \ac{NVM} when energy is back, so executions resume close to the point of energy failure rather than performing a complete reboot.

The price to pay to dump application and system state on \ac{NVM} is \emph{enormous}, especially in \emph{energy}.
The corresponding overhead may reach up to 350\% the cost of the application processing, mainly due to the use of energy-hungry \ac{NVM} technology~\cite{ratchet}.
Using persistent state to cross energy failures has further implications.
When using \ac{FRAM} as \ac{NVM}, for example, wait cycles may be necessary to synchronize read/write operations with the \ac{MCU}, further increasing energy consumption~\cite{msp430fr5969}. 
Finally, if the system employs stateful peripherals, their state  is to be retained across energy failures too~\cite{sytare,restop,samoyed,karma}.
This increases the size of persistent state that must include information that may not be reflected in the system's main memory, adding to the energy overhead.

\fakepar{Trading energy for output quality} Our work starts from the observation that a number of \ac{IOT} applications, as in smart health~\cite{solanas2014smart}, ambient intelligence~\cite{cook2009ambient}, and environment monitoring~\cite{othman2012wireless} \emph{inherently tolerate approximate results}.
This stems from the nature of data processing in these applications, including computer vision, machine learning, signal processing, and pattern identification.
These algorithms offer probabilistic guarantees in the first place and are resilient to data errors, for example, due to sensor inaccuracies~\cite{mittal2016survey}.

Modern \ac{NVM} technology, such as \ac{STT-MRAM}, is a stepping stone to take advantage of these characteristics.
We can pilot \ac{STT-MRAM} chips during write operations to reduce energy consumption, specifically, by applying a variable current intensity.
Due to \textit{stochastic switching}~\cite{Devolder-2008} of \ac{STT-MRAM} cells, further described in \secref{sec:related}, lower current intensity applied during write operations causes \emph{write errors} that are randomly distributed.
Write errors inevitably affect the \emph{quality of the output} the system eventually produces.

Using the knob provided by controlling energy consumption of \ac{STT-MRAM} at the cost of write errors, here we seek to answer a fundamental question, as we articulate in \secref{sec:problem}:
\begin{quote}
  \emph{What are the applications, system settings, and hardware platforms where the trade-off between energy savings and reduction in output quality is favorable?}
\end{quote}
Our technique resonates with approximate computing techniques~\cite{mittal2016survey} aimed at \textit{introducing controlled errors to save resources}. 
What we do here is fundamentally different, however.
Application requirements normally dictate rigid lower bounds on output quality~\cite{han2013approximate}.
Approximate computing techniques aim at ensuring this minimum quality by saving the greatest resources, for example, energy.
In our case, the problem is reversed.
Applications may not impose minimum accuracy requirements and be satisfied with whatever is attainable, whereas \emph{a strict upper bound on energy consumption} exists due to the finite energy buffers.
The notion of output quality, however, the way it is measured, and the definition of acceptable bounds are inherently application-specific.
This is not only germane to our work, but a general characteristic of \emph{approximate computing}~\cite{mittal2016survey, han2013approximate}.

\fakepar{Results and take-aways} Gathering experimental data to answer the question is a challenge per se.
MCUs with integrated \ac{STT-MRAM} are commercially unavailable, forcing system designers to build ad-hoc integrations of off-the-shelf MCUs and \ac{STT-MRAM} chips.
The extreme variety of \acp{MCU} employed for battery-less \ac{IOT} devices depending on computing power, energy consumption, and form factor would also make this a one-off exercise.
Even if concrete hardware was at disposal, the number of different benchmarks, system configurations, and inputs required to build a solid experimental foundation would render running tests on real hardware an endless effort.

To address this issue, we design an experimental framework, described in~\secref{sec:methods}, by combining state-of-the-art \ac{MCU} emulators, accurate \ac{NVM} estimation tools, carefully-selected profiling on real hardware, and executions on mainstream computing machines.
Based on existing literature and the widespread use of these instruments~\cite{batteryless-future,eriksson2009cooja,dong2012nvsim}, the energy figures we obtain, along with the corresponding estimation of the output quality, are arguably on par with real hardware.

Based on $1.9+$ trillion experimental data points, we draw \emph{three fundamental observations}, further discussed in~\secref{sec:results}:
\begin{enumerate}
\item the trade-off between energy saving and reduction of output quality is \emph{program-specific}: we test benchmarks where we achieve up to 50\% reduction in energy consumption with negligible effects on output quality, as opposed to settings where a minimal energy gain causes drastic drops in output quality.
\item the same trade-off is a function of a \emph{platform's specific compute efficiency and power figures}: in the same setting as the previous point, changing the target platform halves the reduction in energy consumption. 
\item \emph{data encoding} and \emph{input size} impact a program's resilience to errors: in the case of neural network benchmarks, as an example, non-quantized networks quickly degrade the output quality as even a single bit error may have disastrous effects.
\end{enumerate}
We end the paper in~\secref{sec:end} with brief concluding remarks.


\section{Related Work and Background}
\label{sec:related}

Our work intersects different areas, including energy harvesting, low-power embedded computing, and hardware architectures.
We briefly survey existing literature relevant to the following discussion and contrast that with our own work.

\subsection{Intermittent Computing}


Operations to persist application, system, and peripheral states entirely represent \emph{an overhead} since they consume energy exclusively to let a program cross energy failures, yet without contributing to making progress in the application logic.
Taming this overhead is the primary focus of a major fraction of existing works in intermittent computing.

Some solutions employ a form of checkpointing~\cite{mementos,Hibernus,Hibernus++,chinchilla,HarvOS,DICE,ratchet}.
This consists in replicating the application state on \ac{NVM} at specific points in the code, where it is retrieved back once the system resumes with sufficient energy.
Systems such as Hibernus~\cite{Hibernus,Hibernus++} operate based on interrupts fired from a hardware device that prompts the application to take a checkpoint, for example, whenever the energy level falls below a threshold.
Differently, systems exist that place function calls in application code to proactively checkpoint~\cite{mementos,HarvOS,ratchet,chinchilla}.
The specific placement is a function of program structure and energy patterns.

Other approaches offer abstractions that programmers use to define and manage persistent state~\cite{dino,alpaca,chain,coala,Ink}.
Most of these employ a \emph{task} abstraction to structure the code.
A task is a single computational unit with defined inputs and outputs, which runs with \emph{transactional semantics} and commits outputs on \ac{NVM} upon completion.
Albeit requiring programmers to re-factor existing code, tasks allow one to precisely map the energy consumption of different code segments to the available hardware, for example, when using multiple capacitors~\cite{colin2018reconfigurable}, as we illustrate next.

Our work is orthogonal to most existing literature in intermittent computing.
Saving energy by controlling the energy invested in \ac{NVM} writes applies to both checkpoint and task-based systems, regardless of whether the data represents application, system, or peripheral states.
The key point is only to apply this technique to data that is error resilient: as we elaborate next, we cannot afford any error when persisting program control data, whereas data encoding images, audio, or signals is inherently subject to errors.   



\subsection{Data Errors and Non-volatile Writes}

As hinted above, \ac{IOT} applications in diverse domains inherently tolerate approximate results.
The underlying data pipelines are resilient to errors by design.
For instance, \acp{NN} used for image classification generally work properly also in the presence of noisy input images, as they provide probabilistic estimates in the first place. 
The corresponding data pipelines also often provide various parameters, such as image resolution, that can be tuned to trade resource consumption against output quality.
These features become a stepping stone to design approximate computing techniques~\cite{mittal2016survey} aimed at \textit{purposely introducing data errors to save resources}.

\acp{NVM} represent an alternative to \ac{SRAM} technology with higher density and speed, lower energy consumption, and the ability to persist data.
As discussed in the International Technology Roadmap for Semiconductors report~\cite{itrs}, due to higher energy efficiency compared to other \acp{NVM}, \ac{STT-MRAM} technology is an asset, especially for resource-constrained systems like energy-harvesting devices.
However, \ac{STT-MRAM} suffers from \textit{stochastic switching}~\cite{Devolder-2008}: during write operations, a memory cell may fail to commute to the new input value, thus causing a \textit{write error}. 
This phenomenon is shown to depend on the current intensity applied during writes~\cite{Cast2020}; the lower the intensity, the higher is the probability of an error.
Being the write errors stochastic in nature, errors are introduced in a random way, yielding a ``salt and pepper'' pattern in saved data.

Efforts exist where stochastic switching of \ac{STT-MRAM} is used as a knob to gain resources when the memory is deployed as cache or scratchpad~\cite{Oboril-VTS-2016,STT-RAM-SPM-DAC-2015,Cast2020,Seyedfaraji2022}. 
To structure the analysis, these works define a limited set of \emph{\acp{QL}}~\cite{Cast2020}, each one of them characterized by a specific write current intensity and thus by a defined probability of write errors per single memory bit write.
We do the same, as further explained in \secref{sec:methods}, and define five distinct \acp{QL} to configure the system ahead of running tests and discuss the corresponding results.
In this work, the setting of quality level is the tunable knob to trade energy consumption for output quality.

Unlike existing literature, however, here we explore the use of \ac{STT-MRAM} with current scaling in an intermittent system, using  \ac{STT-MRAM} to persist states.
Current scaling resonates with near-threshold computing~\cite{dreslinski2010near}, which however controls supply voltage and not current, and is most often applied to computing units and not memories, especially low-power ones. 
Overall, we  may regard this technique as an instance of approximate intermittent computing~\cite{bambusi2022case} working close to the hardware and not requiring changes to algorithms. 


\section{Problem and Objectives}
\label{sec:problem}

We describe the essential features of the platforms we target and their configuration, articulate the benefits we obtain from controlling the \acp{QL} in \ac{STT-MRAM} writes, and weigh them against the cost of reducing the output quality.
We conclude by formulating the problem we tackle and the corresponding objectives.

\subsection{Platforms and Configurations}

\begin{figure}[tb]
    \includegraphics[width=0.7\linewidth]{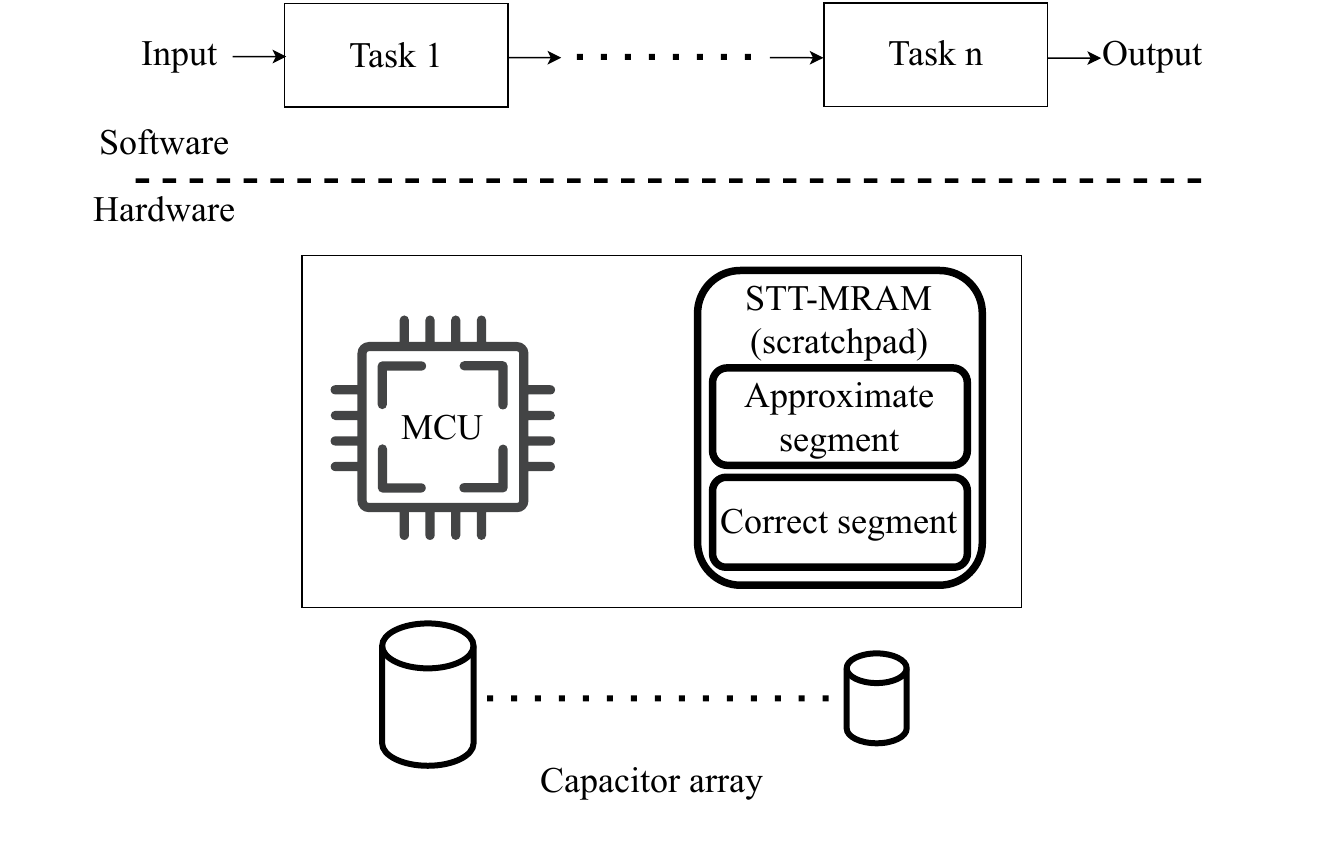}
    \caption{Abstract representation of target platform.}
    \label{fig:hwsw}
\end{figure}

We consider a platform composed of a resource-constrained \ac{MCU} using \ac{STT-MRAM} as a non-volatile scratchpad memory, as abstractly depicted in \figref{fig:hwsw}.
Concrete examples of target \acp{MCU}  are the MSP430 or Cortex M series, as seen in real deployments of battery-less \ac{IOT} systems~\cite{water-deployment-microbial-fuel-cell,tethys,soil-termoelectric,sensys20deployment,denby2023kodan}.
To achieve energy efficiency, these MCUs are extremely primitive in hardware features and lack, for example, functionality such as \ac{DMA}.

Using the \ac{STT-MRAM} as a scratchpad memory, while tuning the current intensity during write operations, requires extreme care. 
If errors are introduced on arbitrary program data, chances are that errors may also corrupt data affecting the program's execution flow.
This would be the case, for example, if errors affected control variables in a program, say causing unwanted jumps in the code.
We exclude this kind of data from write operations where we tune the current intensity, and tune write currents only with data that is potentially resilient to errors, as in the case of imagery or sensor data. 
This may be achieved on the same \ac{STT-MRAM} chip: it is possible to divide the memory into segments, each with a different write current~\cite{Cast2020}.
We reserve one segment to control data with a (probabilistic) guarantee that no errors are introduced.

In line with state-of-the-art intermittent systems, we consider task-based programming, capacitors as energy buffers, and a multi-capacitor setup~\cite{colin2018reconfigurable}, as shown in \figref{fig:hwsw}.
The latter allows designers to optimize capacitor size based on the energy demands of the tasks in a program.
A large capacitor is used to store the energy required for the execution of energy-hungry tasks, whereas smaller capacitors are used for tasks with lower energy demands. 
This improves the general performance because it better matches energy availability and demand. 
Large capacitors are characterized by slower recharge times and high leakage.
They are only used when necessary, that is, to support energy-hungry tasks. 
If tasks can run to completion with smaller energy budgets, they rely on smaller capacitors with shorter recharge times and reduced leakage~\cite{colin2018reconfigurable,flicker}.

\subsection{Benefits}

\begin{figure}[t]
    \includegraphics[width=\linewidth]{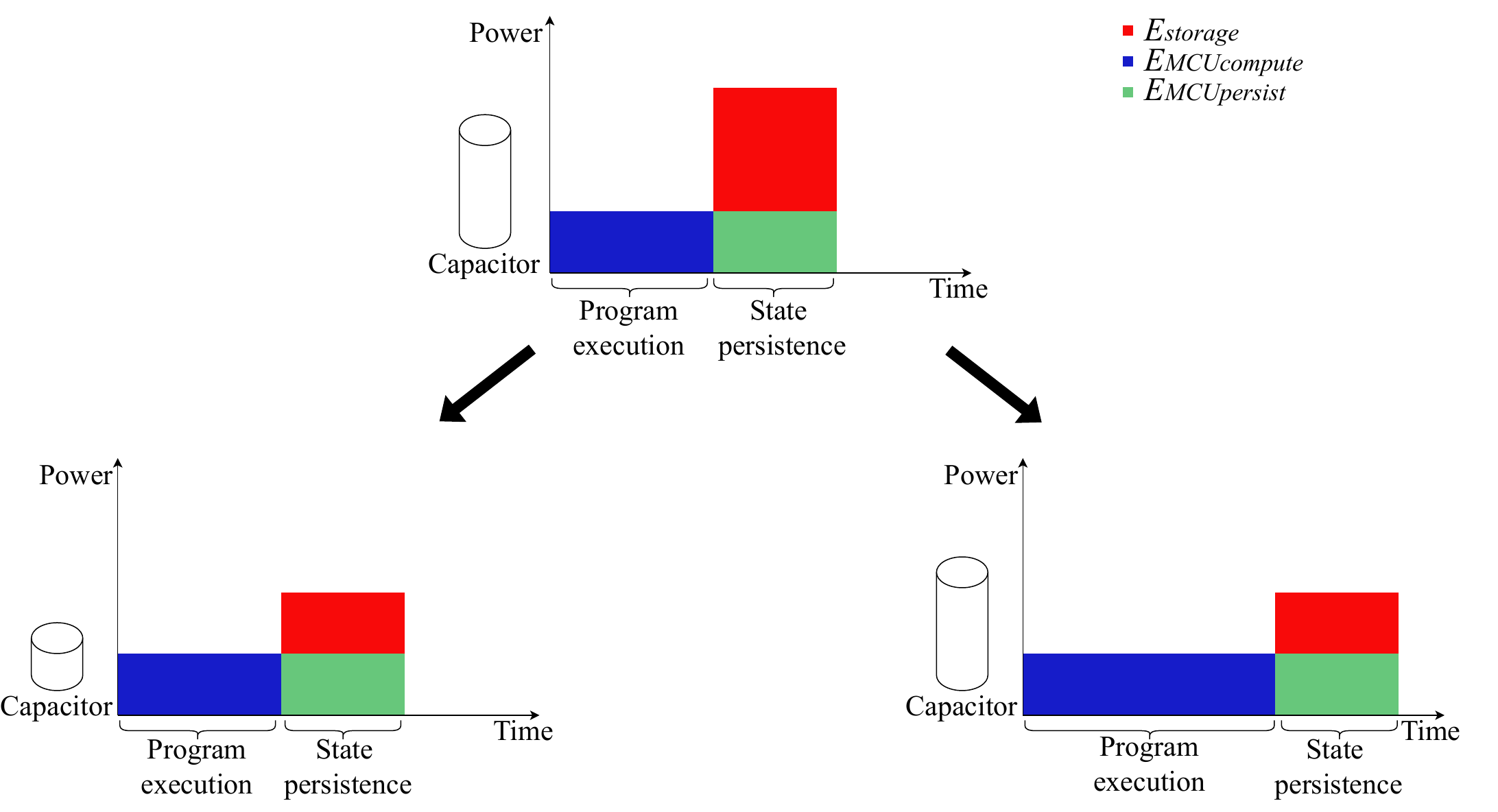}
    \caption{At the top, we have an example of task execution, while at the bottom, we explore two methods to enhance the system's performance. In the bottom-left, we consider reducing the capacitor dimension, and in the bottom-right, we consider prolonging the code execution. }
    \label{fig:overhead}
\end{figure}

Intermittent systems persist the state on \ac{NVM} to cross energy failures.
The upper side of \figref{fig:overhead} qualitatively illustrates the energy patterns during an example task execution.
The \ac{MCU} consumes a variable portion of the available energy for executing code, represented as the blue area in \figref{fig:overhead}, thus making progress in the application logic.
In anticipation of an energy failure, application, system, and peripheral state are persisted on \ac{NVM} when the task completes. 
Because of the lack of features such as \ac{DMA}, the \ac{MCU} remains active performing the data transfer. 
During these operations, the energy consumed by the \ac{MCU} adds to the energy spent in \ac{NVM} writes.

Persisting the state thus subtracts energy that could be rather invested in useful processing and hence represents an overhead that should be minimized.
This overhead corresponds to the red area in \figref{fig:overhead}, which is a function of current intensity, write latency, and operating voltage of the \ac{STT-MRAM} chip.
All other memory operating settings staying the same, we can shrink the red area by piloting the current intensity.

One may invest the energy we save to improve the system's performance in two ways: 
\begin{enumerate}
\item Using a smaller capacitor while keeping the same energy invested in executing code, as shown in the bottom left \figref{fig:overhead}. If the red area shrinks, the total amount of energy required to complete a task, represented as the sum of the red and blue areas, also diminishes.
As smaller capacitors recharge faster and enjoy reduced leakage, the system throughput increases. 
\item Shifting the energy gained by shrinking the red area to the blue area, as shown in the bottom right of \figref{fig:overhead}, that is, to code execution.
  Keeping capacitors fixed, gains in the energy required for \ac{STT-MRAM} writes may be invested in making a task accomplish additional useful work; by doing so, a program may finish the workload sooner, again increasing the system's throughput.
\end{enumerate}

\subsection{Drawbacks}

Tuning the \ac{STT-MRAM} write current comes with a major drawback, write errors may occur that affect the quality of the output, possibly rendering the system useless.

\begin{figure}[t]
    \includegraphics[width=0.75\linewidth]{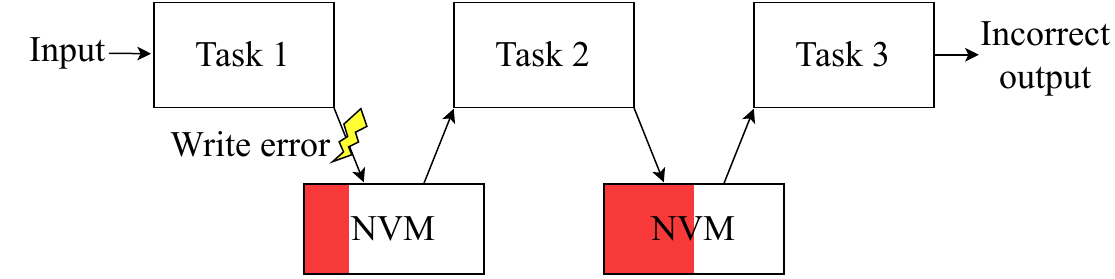}
    \caption{Error snowball effect, the red area represents corrupted data.}
    \label{fig:chain}
\end{figure}

Fully understanding the impact of these errors is non-trivial.
Task-based programming is a form of data-flow programming~\cite{halbwachs1991synchronous}.
A single write error occurrence when persisting the output of a task early in the pipeline may propagate as an incorrect value throughout multiple tasks downstream, each taking as input the output of the previous one.
This would create a snowball effect, intuitively described in \figref{fig:chain}, ultimately heavily affecting the final output.

Two aspects are, therefore, to be taken into account: \textit{i)} a single write error does not necessarily result in only one erroneous value in the final output, and \textit{ii)} not all write errors bear the same impact on the final output.
This is especially true when dealing with bit flips, as it happens on \ac{STT-MRAM}; trivially, if the bit switch occurs on the most relevant bit of a value, the effect is certainly more disruptive than if it occurs on the least significant one.

These observations lead us to realize that providing a statistically sound assessment of the effects of tuning write currents of \ac{STT-MRAM} output quality requires running a substantial number of experiments.
Each experiment may have errors that affect the execution differently, emphasizing the need for a comprehensive evaluation.

\subsection{Problem}

The problem we address is therefore to \textit{quantify the trade-off between the energy savings enabled by tuning write currents of \ac{STT-MRAM} and the reduction in the output quality}.
This is a multi-faceted problem, because \emph{i)} the factors ultimately determining the energy savings are multiple, and \emph{ii)} the notion of output quality is application-specific.

As for \emph{i)}, the energy consumed by a battery-less \ac{IOT} device to execute a given task given a sufficient energy budget is
\begin{equation}
  E_{total} = E_{MCUcompute} + E_{MCUpersist} + E_{storage}
\label{eq:etotal}
\end{equation}
where $E_{MCUcompute}$ is the energy spent by the MCU in useful computations, shown in blue in \figref{fig:overhead}, $E_{MCUpersist}$ is the energy spent by the MCU while the \ac{STT-MRAM} writes data, shown in green in \figref{fig:overhead}, and $E_{storage}$ is the energy spent by the \ac{STT-MRAM} when writing data, shown in red in \figref{fig:overhead}.

Each term in \eqref{eq:etotal} may be rewritten, piece by piece, depending on the characteristics of the hardware at hand, leading to
\begin{equation}
  \begin{split}
  E_{total} &= E_{cycle} \cdot \#\mathit{cycles} + E_{cycle} \cdot \#\mathit{memAccesses}  \cdot  + E_{memWrite} \cdot \#\mathit{memAccesses} \\ &=E_{cycle} \cdot \#\mathit{cycles} + (E_{cycle}  + E_{memWrite}) \cdot \#\mathit{memAccesses}
  \end{split}
\label{eq:etotal-hardware}
\end{equation}
where $E_{cycle}$ is the MCU energy for a single compute cycle, $\#\mathit{cycles}$ is the number of MCU cycles to complete the tasks chosen for execution, $\#\mathit{memAccesses}$ is the number of memory accesses during state persistence operations, and $E_{memWrite}$ is the write energy for a single memory access.

Given a task $T$ of a program, we make explicit how the quantities of \eqref{eq:etotal-hardware} depend on multiple software and hardware factors as
\begin{equation}
  \begin{split}
    E_{cycle}  = f(\mathit{MCU_{activePower}, \mathit{MCU_{frequency}}}),\\
    \#\mathit{cycles} = f(MCU_{ISA}, T), \\
    \#\mathit{memAccesses} = f(T),\\
    E_{memWrite} = f(\mathit{QL}).
\label{eq:params}
  \end{split}
\end{equation}
where $MCU_{activePower}$ and $MCU_{ISA}$ capture the specific power figures and instruction set architecture of the chosen \ac{MCU}, $\mathit{MCU_{frequency}}$ is the \ac{MCU} operating frequency, and $\mathit{QL}$ is the selected \acl{QL} for \ac{STT-MRAM} writes, corresponding to a specific current setting, as in \secref{sec:related}.

To understand the trade-offs at hand, we must quantify the energy gains as the \ac{QL} changes, depending on the executing task and target \ac{MCU}.
The latter two are the key experimental parameters driving the measurements we take, based on the methodology explained next.


\section{Methodology}
\label{sec:methods}

We describe the evaluation methodology we use to produce the quantitative results presented in \secref{sec:results}.
In the following, \secref{sec:benchmarks} introduces the list of selected benchmarks, the metrics we consider, and the target hardware platforms, \secref{sec:framework} describes the experimental framework, and finally \secref{sec:setup} explains the  concrete setup.

\subsection{Benchmarks, Metrics, and MCUs}
\label{sec:benchmarks}

There is arguably no agreed-upon set of benchmarks for the evaluation of intermittent computing systems. 
\tabref{tab:bench} lists the benchmark applications we consider.
The benchmarks span various domains where battery-less \ac{IOT} systems are typically employed~\cite{water-deployment-microbial-fuel-cell,tethys,soil-termoelectric,sensys20deployment,denby2023kodan}, allowing us to provide a robust understanding of the applicability and advantages of our technique under different workloads.
This variety is instrumental in evaluating the trade-off between energy savings and output quality as parameter $T$ in \eqref{eq:params} varies. 

We select several applications from the Mibench2~\cite{MiBench2} benchmark suite, generally employed in the literature~\cite{ratchet,mementos,Hibernus,Hibernus++,chinchilla,HarvOS,DICE, dino,alpaca,chain,coala,Ink} on intermittent computing, including \texttt{FFT} signal processing, \texttt{PicoJPEG} image encoder, and \texttt{Susan} edge detection.
We also consider \acp{NN} because of the growing interest in intermittent systems~\cite{gobieski2019intelligence,islam2020zygarde}.
\acp{NN} exhibit inherent error resilience, making them amenable to approximation.
As representative of this class of workloads, we select six benchmarks provided by ST Microelectronics in their ModelZoo suite~\cite{stm32ai-modelzoo}  that perform image classification: including \texttt{MobileNetV1\_0.25\_96}, \texttt{MobileNetV1\_0.25\_224}, \texttt{FDMobileNet\_128}, \texttt{FDMobileNet\_224}, \texttt{SqueezeNetV1.0} and \texttt{STMnist}. 
The value \_96,\_128 and \_224 for FDMobileNet and MobileNetV1 indicate the input size. 
Depending on this value the \ac{NN} accept RGB images of size 96x96, 128x128, or 224x224.
In the two MobileNetV1 we considered \_0.25 indicates the alpha value of the network, which is the scaling on the number of channels compared to the original MobileNet implementation.
All \acp{NN} are 8-bit quantized. 
Orthogonal to the benchmarks above, we implement an ad-hoc micro-benchmark, called \texttt{OnlyWrites}, that only performs the write of a byte array on \ac{NVM} with no computing workload.

\begin{table}[t]
\centering \begin{tabular}{|@{\phantom{.}}l@{\phantom{.}}|@{\phantom{.}}c@{\phantom{.}}|@{\phantom{.}}c@{\phantom{.}}|@{\phantom{.}}c@{\phantom{.}}|@{\phantom{.}}c@{\phantom{.}}|}
    \hline
    \textbf{Benchmark}&\textbf{Domain}&\textbf{Suite}&\textbf{Multiple tasks}&\textbf{Quality metric}\\
    \hline
    \hline
    {\tt FFT}&Signal processing & Mibench2 & No & \acs{ARE}\\
    {\tt PicoJpeg}&Image processing & Mibench2 & No & \acs{RMSE}\\
    {\tt Susan}&Image processing & Mibench2 & Yes & Precision and recall\\
    {\tt MobileNetV1\_0.25\_96}  & Image classification & ST ModelZoo  & Yes &Accuracy\\
    {\tt MobileNetV1\_0.25\_224} & Image classification & ST ModelZoo & Yes & Accuracy\\
    {\tt FDMobileNet\_128} & Image classification & ST ModelZoo & Yes & Accuracy \\
    {\tt FDMobileNet\_224} & Image classification & ST ModelZoo & Yes & Accuracy \\
    {\tt SqueezeNetV1.0} & Image classification & ST ModelZoo & Yes & Accuracy \\
    {\tt STMnist} & Image classification & ST ModelZoo & Yes & Accuracy \\
    {\tt OnlyWrites} &Micro-benchmark & -- & No & -- \\
  \hline
  \end{tabular}
  \caption{Benchmarks for experimental evaluation.}
  \label{tab:bench}
\end{table}

Depending on code structure and energy demands, a complete run of a benchmark would be implemented as a single task, thus immediately producing an output and incurring in a single state persistence operation at the end, or as a pipeline of tasks, thus incurring in multiple state persistence operations to save intermediate results and cross energy failures. 
\texttt{FFT}, \texttt{PicoJPEG}, and \texttt{OnlyWrite} are implemented as a single task and only the final output is possibly affected by errors.
Susan and all \acp{NN} belong to the latter category, as reported in \tabref{tab:bench}. 


We select a benchmark-specific quality metric to evaluate the impact of write errors on output quality.
We talk about \emph{\ac{QoR}} of an approximated execution compared to a correct one, that is, one where write errors in \ac{NVM} operations are (probabilistically) guaranteed not to happen. 
For \texttt{PicoJPEG}, we employ \ac{RMSE} to measure the average magnitude of the differences between images output of approximate and correct executions~\cite{yazdanbakhsh2016axbench}. 
For \texttt{FFT}, we select the \ac{ARE}, quantifying the average percentage difference between the approximated and correct signals generated as output~\cite{yazdanbakhsh2016axbench}. 
Precision and recall allow us to assess whether the pixels identified as edges in a correct execution of \texttt{Susan} are also identified as edges in an approximate one. 
For the \acp{NN} we employ an ordinary accuracy metric~\cite{stm32ai-modelzoo,mohri2018foundations}, which measures the ratio of correctly classified instances over the total tested instances.
As a micro-benchmark with no computing stage, no quality metric is applicable to \texttt{OnlyWrites}.

\begin{table}[t]
  \begin{tabular}{|l|c|c|c|c|}
    \hline
    \multirow{2}{*}{\textbf{\acs{MCU}}}& \multirow{2}{*}{\textbf{Datasheet}}&\textbf{Clock speed}&\textbf{Main memory} & \textbf{$\mathbf{MCU_{activePower}}$}\\
    & &$(Mhz)$&$(KiB)$ & $(uW/Mhz)$\\
    \hline
    \hline
    MSP430G & \cite{ti-msp430g2453-q1} & 16 & 0.512 & 503\\
    MSP430L & \cite{ti-msp430l092} & 4 & 2 & 58.5 \\
    MSP430S & \cite{singhal20158} &16 & 8 & 28.3\\
    Cortex M0 & \cite{arm-cortex-m0} & 40 & 32 &12.5\\
    Cortex M33 & \cite{arm-cortex-m33} & 160 & 768 &12.0\\
    Cortex M4 & \cite{arm-cortex-m4} &80 &128 &32.82\\
    Cortex M7 & \cite{arm-cortex-m7} &480 &  1024 &58.5\\
  \hline
  \end{tabular}
  \caption{Selected \acsp{MCU}.}
  \label{tab:platforms}
\end{table}

We run experiments on an extremely diverse set of computing units, as reported in \tabref{tab:platforms}.
The selection of such a broad range of \acp{MCU} enables us to explore how the $\mathit{MCU_{activePower}}$ and $\mathit{MCU_{ISA}}$ in \eqref{eq:params} impact the trade-off we investigate, depending on the unique characteristics of each \ac{MCU}.
Note that $\mathit{MCU_{activePower}}$ refers to the power consumption of the \ac{MCU} while executing code from main memory; all other peripherals, such as debugging facilities, are disabled except the computing unit and main memory. 

We select the MSP430 \ac{MCU} series for their wide use in intermittent computing~\cite{batteryless-future,flicker,sensys20deployment}.
They are characterized by low energy consumption and cost-effectiveness, yet offer very limited memory and computing capabilities that generally prevent one from running performance-demanding programs.
Cortex M \acp{MCU}, instead, offer sufficient memory and computing power to run sophisticated programs, such as \acp{NN}, and still their energy figures are often on par or even lower than their MSP430 counterparts, enabling their use in intermittent computing systems~\cite{ratchet} too.

\subsection{Framework}
\label{sec:framework}

\begin{figure}[tb]
    \includegraphics[width=\linewidth]{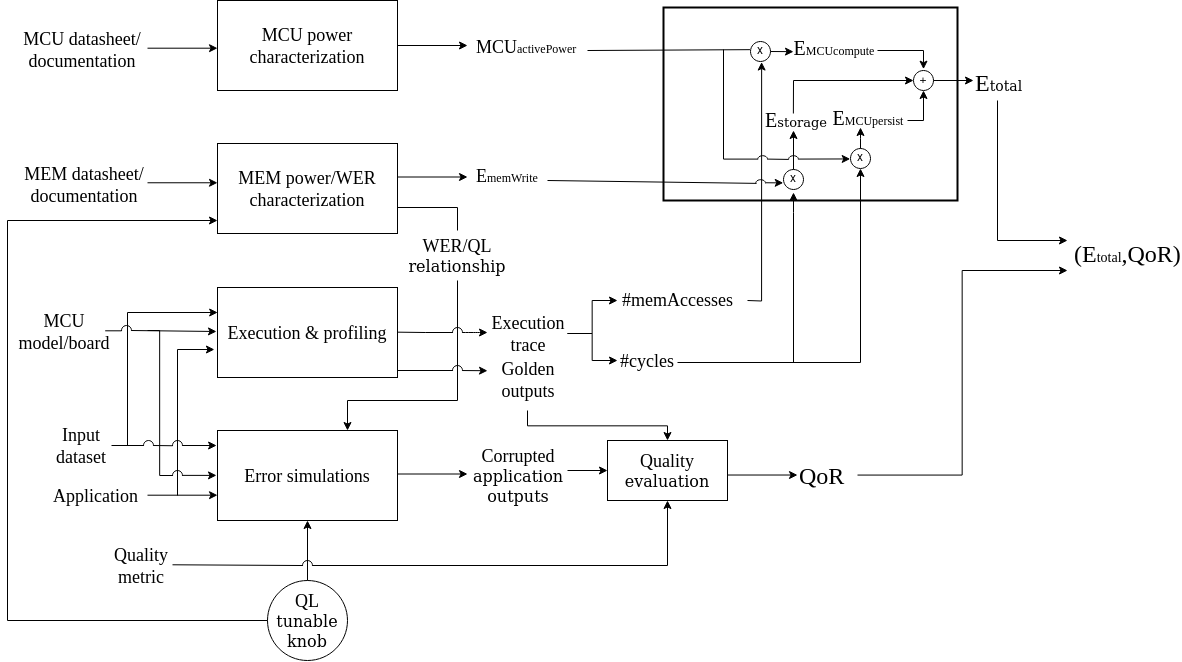}
    \caption{Experimental framework.}
    \label{fig:Framework}
\end{figure}

Commercial \acp{MCU} do not integrate \ac{STT-MRAM} chips yet.
Even if such integration was available, however, the number of different benchmarks, system configurations, and inputs we test would make employing real hardware impractical and extremely time consuming.

\fakepar{Conceptual design} We combine state-of-the-art \ac{MCU} emulators, accurate \ac{NVM} estimation tools, carefully-selected profiling on real hardware, and executions on mainstream computing machines, as shown in \figref{fig:Framework}.
Given a benchmark and a target hardware platform, the framework allows us to measure the energy consumption of the hardware and the corresponding \ac{QoR}, as defined in \tabref{tab:bench}, while applying the different \acp{QL} to \ac{STT-MRAM} writes.



We measure the system's energy consumption by means of the energy model presented in \secref{sec:problem}.
In fact, the development boards hosting \acp{MCU} generally do not provide power sensors for the specific chip, while they also integrate many other debug and utility components that would not be present in a deployed system; as a consequence, ammeters and the like, when hooked to the board, would provide biased measurements.

We compute the system's energy consumption as the sum of \ac{MCU} and \ac{STT-MRAM} energy.
The \acp{MCU} we consider feature very simple architectures with no dynamic voltage/frequency scaling; their energy consumption is a simple function of power consumption and the number of cycles to execute a workload.
Power consumption is retrieved from an \ac{MCU} datasheet.
To compute the number of cycles, we profile a benchmark's execution to extract a trace of machine instructions and memory accesses.
We perform multiple runs to average out the effects of different inputs.

The model we employ for \ac{STT-MRAM} computes the energy consumption by multiplying the energy consumption of a single bit read and write by the overall number of performed operations.
The energy consumption of read operations is retrieved from the \ac{STT-MRAM} datasheet.
For write operations, energy consumption is specific to each \ac{QL}; we explain next how we obtain this value and the corresponding error probability for a single bit write.
Finally, we extract the number of write/read operations from the execution trace above.

The crucial issue is that the evaluation of the \ac{QoR} of a benchmark at a given \ac{QL} requires write errors to be somehow simulated.
Since write errors are stochastic, as common practice in reliability analysis of faulty digital systems~\cite{Classes2023}, we perform a large number of runs of the considered benchmark on the target architecture.
For each such execution, we compute the selected quality metric by comparing the concrete benchmark output, which may be corrupted by write errors on \ac{NVM}, with the \emph{golden} counterpart, that is, the output we collect during the profiling execution above, where no \ac{NVM} write errors occur.
We use multiple different inputs, as explained next, to factor out any bias due to the specific input, and average the \ac{QoR} over all inputs we test.

The outcome of the framework is, for a given benchmark and \ac{MCU}, the corresponding energy consumption, and resulting \ac{QoR} for a specified \ac{QL}. 
By performing the presented flow for all considered \acp{QL}, it is possible to obtain a mapping between each \ac{QL}, the corresponding energy consumption, and the resulting \ac{QoR}.
This mapping is the basis for the analytical evaluation of the trade-off between energy savings and output quality.




\fakepar{Implementation} 
We use NVSIM~\cite{dong2012nvsim} to characterize the \ac{STT-MRAM} and obtain its energy figures. 
NVSIM is a widely used \ac{NVM} simulator (e.g., in~\cite{Cast2020}) that allows system designers to characterize \ac{STT-MRAM} devices, and therefore to estimate read/write currents and energy consumption for a given chip configuration. 
We borrow the mapping between the energy invested for write operations at a given \ac{QL} and the corresponding \ac{WER} for a single bit from existing characterizations~\cite{Cast2020}.

For MSP430 \acp{MCU}, we profile the benchmark executions with MSPSim~\cite{eriksson2009cooja}, a widely-used cycle-accurate MSP430 emulator.
MSPSim provides functionality to extract the execution trace.
On the other hand, we use real development kits from ST Microelectronics to profile the benchmarks on top of Cortex M \acp{MCU}. 
We use hardware counters and on-board debugging facilities to count the number of executed cycles.


As hinted earlier, a solid evaluation of the \ac{QoR} for a benchmark at different \ac{QL} requires a sufficiently large number of runs, which would be unfeasible on real hardware.
We define an application-level error injection approach to this end.
Being \ac{STT-MRAM} operations directly mapped to variable writes, we instrument the benchmark source code with an error injection facility corrupting the bits during each variable assignment with the set \ac{WER} probability.
As this error injection strategy is independent of the hardware running the benchmark, as long as the data encoding is the same, it is possible to execute error simulations on any high-ended workstations and yet obtain the same results of the target \ac{MCU}, but with multi-fold reductions of the experiment time.
We instrument \texttt{OnlyWrites}, \texttt{FFT}, \texttt{PicoJpeg}, and \texttt{Susan} in a semi-automated way, as their source code is quite simple.
The \acp{NN} we consider are implemented with TensorFlow; here we integrate the the error simulation facility in a completely automated way by relying on an existing technique designed to achieve an almost identical effect~\cite{Classes2023}.


\subsection{Setup}
\label{sec:setup}

To structure the analysis, we define five distinct \acp{QL}, each with a different current intensity and corresponding \ac{WER}~\cite{Cast2020}.
We call {\tt Q0} the baseline, with almost 100\% probability of bits to switch correctly and hence ensuring correct executions.
This is the setting we use to obtain the \emph{golden} execution.
Other \acp{QL} range from {\tt Q1} to {\tt Q4}, with the latter offering the greatest energy efficiency albeit with the highest \ac{WER}. 

\begin{table}[b]
  \begin{tabular}{|c|c|c|c|}
    \hline
    \textbf{\ac{QL}}&\textbf{\ac{WER}}&\textbf{Set current  $(\mu A)$}&\textbf{Write energy per bit $(\mathrm{p}J)$}\\
    \hline
    \hline
    {\tt Q0}&$10\textsuperscript{-8}$ & 1153 & 167\\
    \hline
    {\tt Q1}&$10\textsuperscript{-6}$ & 865 & 94\\
    \hline
    {\tt Q2}&$10\textsuperscript{-5}$ & 769 & 74\\
    \hline
    {\tt Q3}&$10\textsuperscript{-4}$ & 673 & 57 \\
    \hline
    {\tt Q4}&$10\textsuperscript{-3}$ & 577 & 43 \\
    \hline
\end{tabular}
  \caption{Characterization of the various \acp{QL}.}
  \label{tab:stt}
\end{table}

\tabref{tab:stt} summarizes the performance figures we obtain from NVSim for each \ac{QL} in the case of a 40nm \ac{STT-MRAM}.
If taken alone, these values indicate that even with a {\tt Q1} setting, energy savings in write operations are substantial, amounting to 56.3\% of the energy consumed at {\tt Q0}.
The reduction in energy consumption becomes less pronounced with higher \acp{QL}, with {\tt Q4} consuming 25.7\% of the energy of {\tt Q0}.
Notably, the decrease in energy consumption does not follow a linear pattern, with a more significant reduction at the initial levels.

\begin{table}[t]
\centering
  \begin{tabular}{|l|c|c|}
    \hline
    \textbf{Benchmark} & \textbf{\# Different inputs} & \textbf{\# Executions}\\
    \hline
    \hline
    \textbf{{\tt PicoJPEG}} & 70,000 & 1,750,000\\
    \hline
    \textbf{{\tt FFT}} & 150,000 & 3,750,000\\
    \hline
    \textbf{{\tt Susan}} & 55,000& 1,375,000\\
    \hline
    \textbf{{\tt STMNIST}} & 75,000& 1,500,000\\
    \hline
    \textbf{{\tt MobileNET\_0.25\_96}} & 15,000& 750,000\\
    \hline
    \textbf{{\tt MobileNET\_0.25\_224}} & 15,000& 1250,000\\
    \hline
    \textbf{{\tt FDMobileNET\_128}} & 15,000& 900,000\\
    \hline
    \textbf{{\tt FDMobileNET\_224}} & 15,000& 1250,000\\
    \hline
    \textbf{{\tt Squeezenet V1.1}} & 15,000& 900,000\\
    \hline
  \end{tabular}
  \caption{Number of different inputs and executions for each benchmark.}
  \label{tab:Experiments}
\end{table}

\begin{table}[t]
\centering
  \scriptsize 
  \begin{tabular}{|l|c|c|c|c|c|c|c|}
    \hline
     & \textbf{MSP430G} & \textbf{MSP430L} & \textbf{MSP430S} & \textbf{Cortex M0} & \textbf{Cortex M33} & \textbf{Cortex M4} & \textbf{Cortex M7} \\
    \hline
    \hline
    \textbf{{\tt PicoJPEG}} & 1214 & 218 & 150 & & & & \\
    \hline
    \textbf{{\tt FFT}} & 16871 & 2005 & 995 & & & & \\
    \hline
    \textbf{{\tt Susan}} & 1021 & 152 & 93 & & & & \\
    \hline
    \textbf{{\tt STMNIST}} & & & & 36, 19 & 24, 5.5 & 29, 10.5 & 30, 14 \\
    \hline
    \textbf{{\tt MobileNET\_0.25\_96}} & & & & & 35, 9.3, 5.6  & 81, 32.7, 8 & 85, 32, 10.2 \\
    \hline
    \textbf{{\tt MobileNET\_0.25\_224}} & & & & & 230, 74, 45 & & 506, 142, 59 \\
    \hline
    \textbf{{\tt FDMobileNET\_128}} & & & & & 75, 19.5, 6.2 & 155, 28.7, 10 & 164, 30.4, 11.5\\
    \hline
    \textbf{{\tt FDMobileNET\_224}} & & & & & 184, 60.9, 21.85 & & 461, 95, 38 \\
    \hline
    \textbf{{\tt Squeezenet V1.1}} & & & & & 399, 220, 43 & & 920, 482, 60 \\
    \hline
    \textbf{{\tt OnlyWrites}} & 5.2 & 3.5 & 3.4 & 3.3 & 3.2 & 3.5 & 3.4 \\
    \hline
  \end{tabular}
  \caption{Capacitor sizes (in $uF$) selected for each benchmark and platform. When no value is shown, that benchmark does not run on the given platform.}
  \label{tab:expS}
\end{table}

In \tabref{tab:Experiments} we show for each benchmark the number of inputs and executions we run for our experimental evaluation.
The number of inputs for all the \acp{NN} except STMNIST is the same  since these are all trained in the same dataset, and we use their dataset for the evaluation.
Note that \texttt{OnlyWrites} is not present in the table since we do not conduct an experimental campaign to assess its \ac{QoR} variations given that it has no quality metric, thus we run it only once.

\tabref{tab:expS} shows what benchmark runs on what \ac{MCU}, and using what capacitor(s).
Indeed, not every benchmark can execute on every \ac{MCU} we consider, mainly due to memory limitations.
We generally run benchmarks on the platform that provides best energy efficiency for the given workload, or that was used in real deployments to the same end.
For example, we run benchmarks from MiBench2 on the very resource constrained MSP430 platforms, while we execute \acp{NN} on Cortex M \acp{MCU}.
Of the Cortex M series, the M0 \ac{MCU} is the one with lowest memory resources, allowing us to experiment only with the small {\tt STMNIST} \ac{NN} that requires less than 32KB of main memory.
The Cortex M4 instead can host {\tt Mobilenet\_0.25} with inputs of dimension 96x96, but it cannot host the same model with inputs of size 224x224.
The same applies to {\tt FDMobilenet}.
On the other hand, Cortex M33 and M7 \acp{MCU} can host all \acp{NN} we consider. 

We chose the capacitor values reported in \tabref{tab:expS} size through a mixed analytical and experimental approach~\cite{teg-apps}, as in real deployments~\cite{sensys20deployment,magno2016infinitime}.
In essence, with a benchmark that only includes a single task, the size of the single capacitor must match the peak energy demand of the task to ensure completion in the worst case, that is, when the environment provides no additional energy while the benchmark executes.
In the presence of multiple tasks and using multiple capacitors, the largest capacitor is  dimensioned to ensure completion of the most energy-demanding task in the same worst-case conditions above.
Two additional capacitors of small and medium size are employed to support the execution of short-lived tasks and to provide a halfway measure~\cite{teg-apps,colin2018reconfigurable}.


\section{Results and Take-aways}
\label{sec:results}

We present next quantitative evidence of the trade-off between energy savings and \ac{QoR} reduction, depending on the chosen \ac{QL}.
\tabref{tab:observations} summarizes the key observations we can draw based on $1.9+$~trillion experimental data points, serving as a road-map throughout the rest of this section.

\begin{table}[t]
  \begin{tabular}{|p{340pt}|c|}
    \hline
    \textbf{Observation} & \textbf{Section} \\
    \hline
    \hline
    The trade-off between energy savings and degradation of output quality is benchmark-specific and an experimental evaluation is strictly required, yet the behavior at the extremes of the parameter spectrum is fairly predictable.  & \ref{sec:resultsTradeoff} \\
    \hline
    The trade-off between energy savings and degradation of output quality is also \ac{MCU}-specific, and it becomes favorable when the energy to persist states outweighs the energy to generate the aforementioned state. & \ref{sec:resultsMCU} \\
    \hline
    Quantization impacts \acp{NN} resilience to errors introduced during state persistence operations and actually becomes an asset to limit the degradation of output quality, compared to their non-quantized counterparts. & \ref{sec:quant} \\
    \hline
    \acp{NN} with larger input size are typically more error resilient; conversely, \acp{NN}  with smaller input sizes are more vulnerable to errors precisely in the parameter range that is likely most significant.  & \ref{sec:inputSize} \\
    \hline
  \end{tabular}
  \caption{Key observations and take-aways.}
  \label{tab:observations}
\end{table}

\subsection{Upper Bound in Energy Savings}

The magnitude of overall energy savings enabled by tuning the write current on \ac{STT-MRAM} depends on how large is the memory contribution compared to the total energy figure.
This is because tuning the \ac{QL} affects only the $E\textsubscript{storage}$ factor in \eqref{eq:etotal}.
The other factors in \eqref{eq:etotal} are, instead, a function of the chosen \ac{MCU}.
To understand this in a limit case, we run the \texttt{OnlyWrites} micro-benchmark that only writes a byte array on \ac{NVM} without any further processing.
This means we place ourselves in a case where the blue area in the top part of \figref{fig:overhead} is null.
The memory contribution to the total energy figure is thus maximum, and so is the energy reduction that shaving off energy from memory operations possibly enables.



\begin{figure}[t] 
    \includegraphics[width=0.85\linewidth]{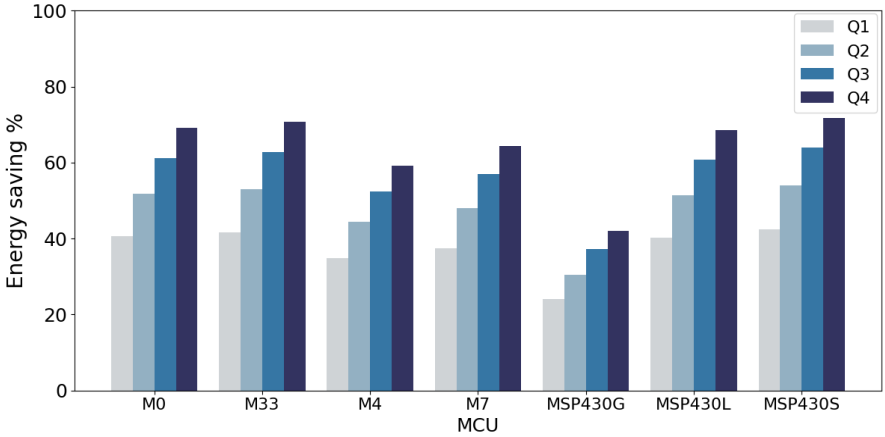}
    \caption{Maximum energy reduction attainable for each \ac{MCU} at different \acp{QL}, compared to the energy consumption at \texttt{Q0}.}
    \label{fig:onlyWrites}
  \end{figure}

\begin{figure}[t]
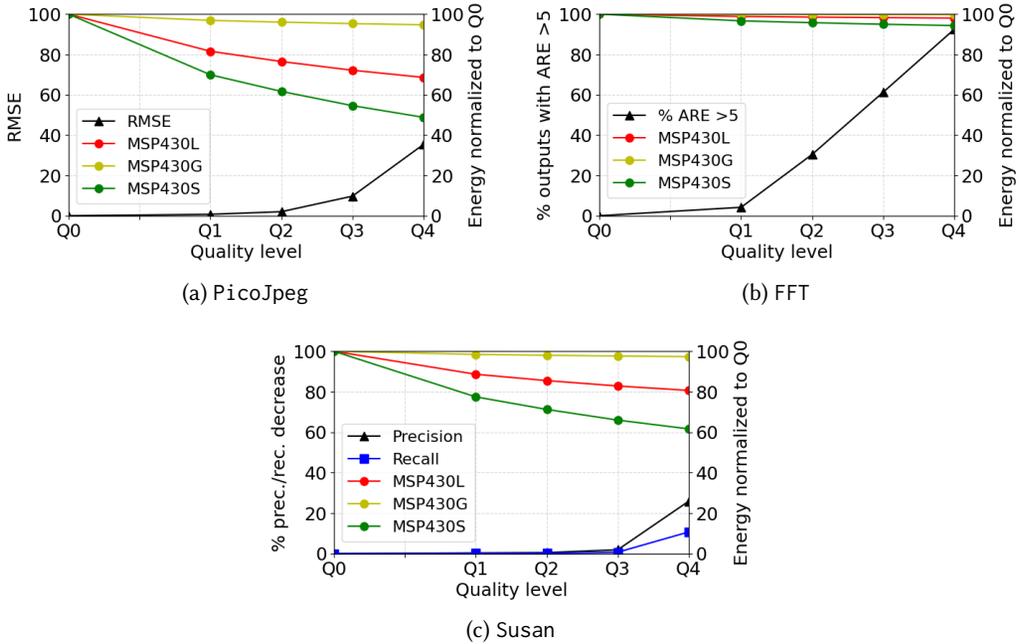

    \centering
    \subfloat[\texttt{PicoJpeg}]{ \includegraphics[width=0.47\textwidth]{Images/Pico.pdf}
    }
    \quad
    \subfloat[\texttt{FFT}]{
        \includegraphics[width=0.47\textwidth]{Images/FFT.pdf}
    }
    \quad
    \subfloat[\texttt{Susan}]{
        \includegraphics[width=0.47\textwidth]{Images/Susan.pdf}
      }
    \caption{Trade-off between energy savings and \ac{QoR} reduction for MSP430 benchmarks.}
    \label{fig:energytradeMSP430}
\end{figure}

\figref{fig:onlyWrites} shows the results.
The energy savings are significant overall, but still span a wide range of absolute values depending on the \ac{MCU}.
This is because no \ac{DMA} is available; therefore the \ac{MCU} is active while transferring data to \ac{NVM}, causing a non-zero $E\textsubscript{MCUpersist}$ contribution in \eqref{eq:etotal}.
The specific value of $E\textsubscript{MCUpersist}$ depends on the \ac{MCU}.
The Cortex M33 emerges as the one with the largest energy savings, with \texttt{Q4} achieving a reduction of 71\% the energy that would be consumed at \texttt{Q0}, whereas the MSP430G is the one producing the least benefits, with a 39\% reduction at \texttt{Q4}.

As expected, \figref{fig:onlyWrites} also shows that the choice of \ac{QL} heavily impacts the energy consumption of \texttt{OnlyWrites}, as memory operations are a prominent factor. 
Consider again the Cortex M33 \ac{MCU} as an example: the 71\% energy saving at \texttt{Q4} becomes a 49\% reduction at \texttt{Q1}. 
This means that, as long as a program is highly resilient to errors, the Cortex M33 is the \ac{MCU} where we can push the \ac{QL} setting the most and reap the greatest benefits in energy, as opposed to other \acp{MCU} where the increase of energy saving with higher \ac{QL} is more limited.

\fakepar{Take-away} Given a specific \ac{MCU} and a \ac{QL} setting, \emph{there exists a maximum limit for the energy savings} we can obtain using our technique, corresponding to when the energy consumed by code execution is minimal.
Out of the platforms we test, the Cortex M33 exhibits the potential for the highest energy savings among Cortex M \acp{MCU}. 
Similarly, for MSP430 \acp{MCU}, we observe that MSP430S stands out with the highest potential energy savings.

\subsection{Energy versus QoR Trade-off}
\label{sec:resultsTradeoff}


  \begin{figure}[t]
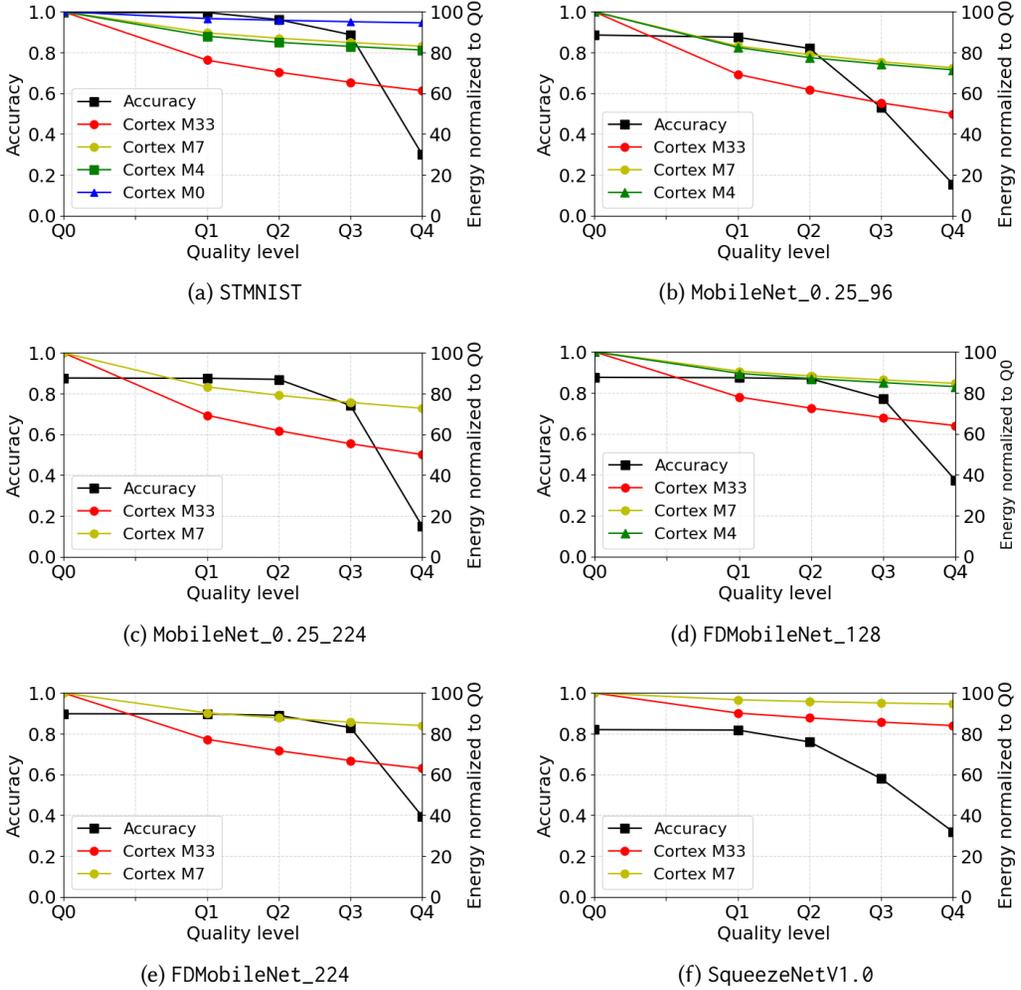

    \centering
  \subfloat[\texttt{STMNIST}]{
        \includegraphics[width=0.47\textwidth]{Images/stmnist.pdf}
    }\quad
        \subfloat[\texttt{MobileNet\_0.25\_96}]{
        \includegraphics[width=0.47\textwidth]{Images/M96.pdf}
    }\quad
        \subfloat[\texttt{MobileNet\_0.25\_224}]{
        \includegraphics[width=0.47\textwidth]{Images/M224.pdf}
    }\quad
        \subfloat[\texttt{FDMobileNet\_128}]{
        \includegraphics[width=0.47\textwidth]{Images/FD128.pdf}
    }\quad
        \subfloat[\texttt{FDMobileNet\_224}]{
        \includegraphics[width=0.47\textwidth]{Images/FD224.pdf}
    }\quad
        \subfloat[\texttt{SqueezeNetV1.0}]{
        \includegraphics[width=0.47\textwidth]{Images/Squeezenet.pdf}
    }
    \caption{Trade-off between energy savings and \ac{QoR} reduction for Cortex M benchmarks.}
    \label{fig:energytradeCortex}
\end{figure}

\figref{fig:energytradeMSP430} and \figref{fig:energytradeCortex} illustrate the trade-off between energy saving and \ac{QoR} degradation for each considered benchmark across benchmarks we run either with MSP430 or Cortex M \acp{MCU}, according to the combinations shown in \tabref{tab:expS}. 
For every benchmark, the plot provides experimental evidence to determine a sweet spot to identify an efficient \ac{QL} setting that balances acceptable \ac{QoR} with maximum energy savings. 

The extent of \ac{QoR} reduction at different \acp{QL} varies for each benchmark. For example, in the case of \texttt{PicoJPEG} in \figref{fig:energytradeMSP430}, the \ac{RMSE} maintains a stable value until the quality level \texttt{Q2}, and increases noticeably only from \texttt{Q3} onwards.
However, still in \figref{fig:energytradeMSP430}, for \texttt{FFT} we notice a drastic decrease in the \ac{QoR} right after \texttt{Q1}. 
In fact, for \texttt{Q2} executions, the outputs may arguably be considered unusable in almost 30\% of cases, showing ARE > 5\%. 
However, even if the \ac{QoR} degradation is benchmark-specific, we note a trend common to all the benchmarks; the \ac{QoR} suffers almost no decrease at \texttt{Q1}, yet a \texttt{Q4} setting makes the results unusable basically in all benchmarks.

\figref{fig:energytradeMSP430} and \figref{fig:energytradeCortex} also demonstrate that the attainable energy saving spans a wide range of absolute values, depending on the benchmark.
This figure ranges from a significant 45\% reduction in \figref{fig:energytradeCortex} for {\tt MobileNet\_0.25\_224} running on Cortex M33 at \texttt{Q3} to a negligible 2\% in \figref{fig:energytradeMSP430} for \texttt{FFT} running on MSP430G even at \texttt{Q4}. 
The reason for this behavior is rooted in the relative balance between the three terms of \eqref{eq:etotal}.
Benchmarks akin to \texttt{FFT} are characterized by large computation phases and relatively small states.
The energy consumed during the code execution dominates other factors, that is, $E\textsubscript{MCUcompute}$ outweighs all other terms in $E\textsubscript{total}$.
Consequently, reducing \ac{STT-MRAM} write energy yields negligible benefits. 
Benchmarks akin to \texttt{MobileNet} feature with large states compared to the computing phase, therefore $E\textsubscript{storage}$ bears a significant impact on $E\textsubscript{total}$.
These benchmarks benefit the most from shaving off energy from state persistence operations.

\fakepar{Take-away} The trade-off between energy saving and \ac{QoR} reduction is specific to a program and the \ac{QL} setting striking the best trade-off is unknown a priori: \emph{an experimental evaluation is strictly necessary}, which may be laborious and time consuming. 
However, in the settings we consider, we note that a write error rate close to 10$^{-3}$, corresponding to \texttt{Q4}, or higher is often degrading the \ac{QoR} to a degree where the results arguably become unacceptable, but provides great energy savings, whereas a write error rate of 10$^{-6}$, corresponding to \texttt{Q1},  or smaller causes almost no degradation in \ac{QoR}, with limited gains in energy consumption.

In essence, the behavior at the extremes of the spectrum is fairly predictable: placing the \ac{STT-MRAM} in a setting where write errors are quite likely produces an excessive degradation of the output, yet provides great energy savings. 
Conversely, configuring the \ac{STT-MRAM} to limit write errors produces a negligible degradation of the \ac{QoR}, with some but not massive energy saving.
In most cases, the experimental evaluation required may be limited to intermediate configurations, where it is more unclear where the trade-off tips over, saving development time and testing efforts.

\subsection{Impact of \ac{MCU} Selection on Energy Savings}
\label{sec:resultsMCU}

\begin{figure}[t] 
\includegraphics[width=0.50\linewidth]{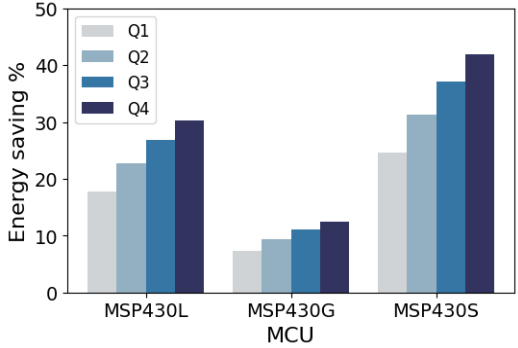}
    \caption{Energy savings for MSP430 \acp{MCU}.}
    \label{fig:AVGONBENCH}
\end{figure}

\begin{figure}[t]
    \centering
    \subfloat[{\tt STMnist} only.]{ \includegraphics[width=0.478\textwidth]{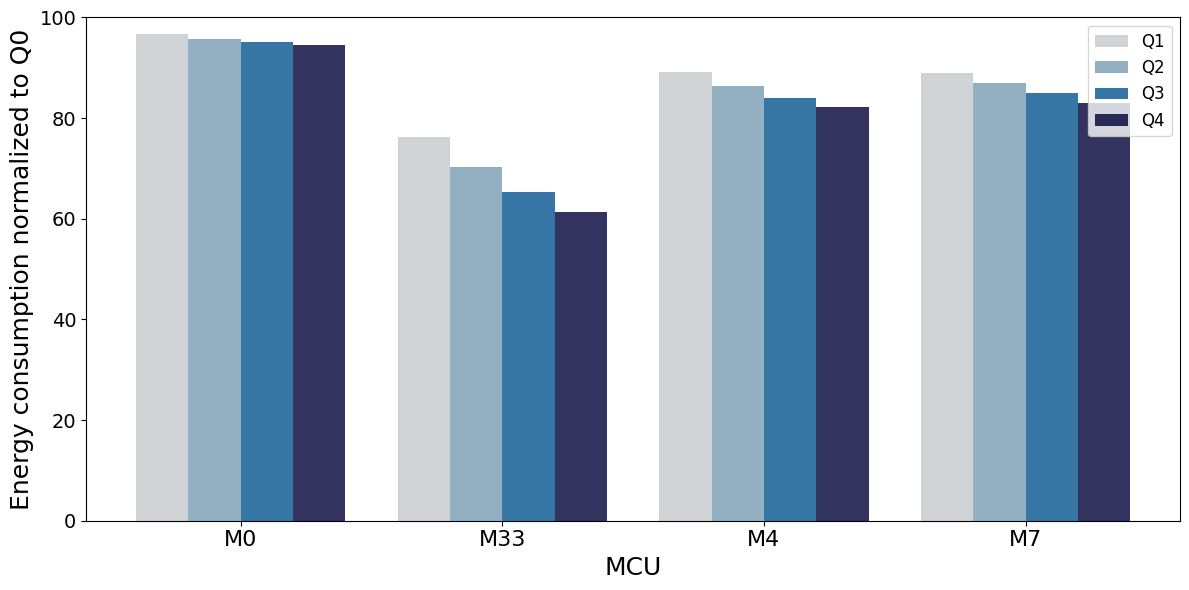}
    }
    \quad
    \subfloat[Average over common benchmarks.]{
        \includegraphics[width=0.462\textwidth]{Images/CortexInCommon.pdf}
    }
    \caption{Energy savings for Cortex M \acp{MCU}.}
    \label{fig:avgCortex}
\end{figure}


\figref{fig:AVGONBENCH} and \figref{fig:avgCortex} show the energy savings enabled by each \ac{MCU} we select, averaged across all the benchmarks we consider.
The extent of achievable energy saving depends on the former, each \ac{MCU} featuring different computing efficiency and power consumption.
For instance, when considering the MSP430 family shown in \figref{fig:AVGONBENCH}, the MSP430S reduces energy consumption by about 40\% with a \texttt{Q4} setting compared to the baseline at \texttt{Q0}.
On the other hand, in the same configurations, the less energy-efficient MSP430G only produces a 10\% reduction in the energy consumption. 

Discussing the results for Cortex M \acp{MCU} requires splitting the data over the benchmarks that can execute on the same subset of \acp{MCU}.
\figref{fig:avgCortex}(a) shows the energy savings for \texttt{STMNIST}, the only \ac{NN} that can run on all Cortex M \acp{MCU}.
The plot confirms the trend; the magnitude of energy savings depends on the \ac{MCU} we consider.
Interestingly, the Cortex M0 is the Cortex M \ac{MCU} where we obtain the least energy savings, even if its active power is lower than that of the Cortex M4 and M7, as reported in \tabref{tab:platforms}.
Its very limited instruction set, which is not optimized for the execution of \acp{NN}, leads to a considerably longer execution time.
In fact, the number of cycles required to execute the inference process on this \ac{MCU} is 20x the one required by the Cortex M7.
Even if active power is limited, the execution stretches in time, causing the device to consume an energy that outweighs the energy invested in \ac{NVM} operations. This makes the reduction of energy spent in the latter operation almost immaterial. 

\figref{fig:avgCortex}(b) shows the energy savings for the Cortex M33, M4 and M7 averaged across the benchmarks they can execute, as per \tabref{tab:expS}. 
The figure confirms the trend seen earlier for \texttt{STMNIST}; the Cortex M33 obtains the best absolute results.
the Cortex M4 and M7 yield similar results, even if the Cortex M7 active power is $\approx$50\% higher than that of the M4.
A similar observation as seen before for the Cortex M0 applies here, but in the opposite way.
The Cortex M7 instruction set is both lager and optimized for \acp{NN} execution, resulting in fewer cycles required to execute the inference step.
In fact, an inference step executed on a Cortex M4 requires $\approx$1.4x the cycles of the Cortex M7.

\fakepar{Take-away}  Scaling \ac{STT-MRAM} write current offers significant energy savings \emph{when the energy needed to persist states outweighs the energy required by the computing phase to generate the aforementioned state}.
The energy spent in computing depends on multiple factors: its duration in time, which is a function of the \ac{MCU} instruction set and frequency setting, and the specific \ac{MCU} active power.
The most favorable case is when the system runs a program with large states produced by short computing phases, for example, as in \texttt{Mobilenet\_0.25\_96}, running on an \ac{MCU} with low active power, as the Cortex M33.

\subsection{Quantization}
\label{sec:quant}

\begin{figure}[t]
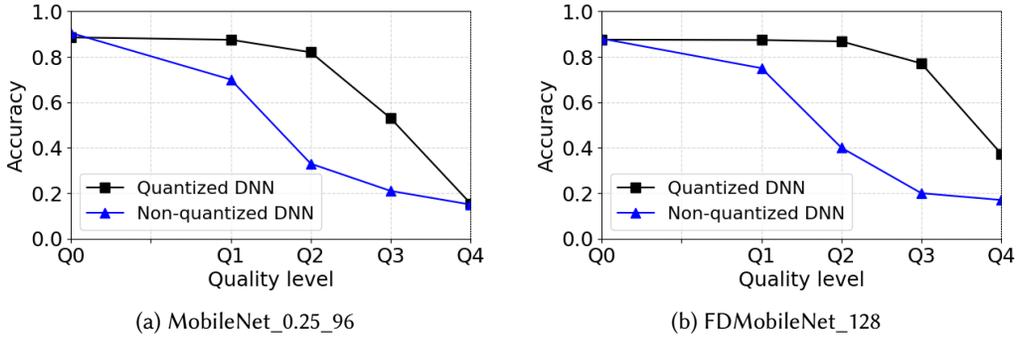

    \centering
    \subfloat[MobileNet\_0.25\_96]{ \includegraphics[width=0.47\textwidth]{Images/Quantization1.pdf}
    }
    \quad
    \subfloat[FDMobileNet\_128]{
        \includegraphics[width=0.47\textwidth]{Images/Quantization2.pdf}
    }
    \caption{Effect of quantization on \acp{NN} error resilience.}
    \label{fig:quant}
\end{figure}

The \acp{NN} used in these experiments are 8-bit quantized, specifically, by converting 32-bit floating-point representations of weights and intermediate values of \acp{NN} into 8-bit integers. 
Quantization is commonly performed when deploying a \ac{NN} onto a resource-constrained platform to reduce the memory requirements and computing demands. In particular, the considered quantization entails. 
This choice is guided by key critical considerations:
\begin{enumerate}
\item Non-quantized \acp{NN} are more demanding in terms of resources, necessitating a minimum of four times the memory compared to their quantized counterparts. In the context of intermittent systems, which inherently face resource constraints, as others did already, we observe that several \acp{NN} chosen for our evaluation would exceed the storage capacity of the selected platforms otherwise.
  This is not just our own observation~\cite{gobieski2019intelligence,islam2020zygarde}.
    \item \acp{NN} with 8-bit integer quantization exhibit better error resilience compared to their non-quantized counterpart, because intermediate values have a restricted numerical range. This property is widely discussed in the investigation on \ac{NN} resiliency~\cite{ranger}, when it is observed that restricting the numerical range of the \ac{NN} intermediate values reduces the deviation caused by bit flips to a level that can be tolerated.
\end{enumerate}

\figref{fig:quant} compares the results of error simulation for two quantized \acp{NN} compared to their non-quantized counterparts.
Our experiments confirm the error resilience properties of \acp{NN}.
We note that the accuracy of non-quantized networks, even at a \texttt{Q1} setting, is lower than the accuracy achieved at \texttt{Q3} on the quantized network. This trend holds true for other networks we experimented with, demonstrating a consistent behavior across multiple settings.

\fakepar{Take-away}  Quantization of \acp{NN} is necessary not just to meet memory requirements of battery-less \ac{IOT} devices, but it \emph{becomes an asset to limit the degradation of output quality}, compared to their non-quantized counterparts,  when state persistence operations are subject to errors. 

\subsection{Input Size}
\label{sec:inputSize}

\begin{figure}[t]
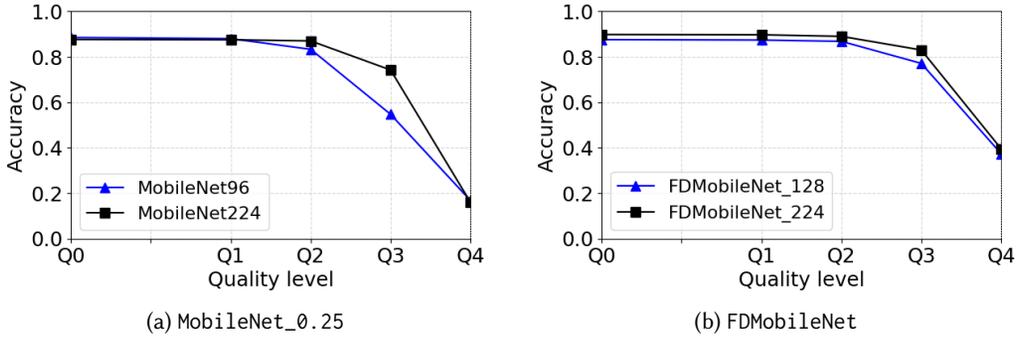

    \centering
    \subfloat[\texttt{MobileNet\_0.25}]{ \includegraphics[width=0.47\textwidth]{Images/inputSize1.pdf}
    }
    \quad
    \subfloat[\texttt{FDMobileNet}]{
        \includegraphics[width=0.47\textwidth]{Images/inputSize2.pdf}
    }
    \caption{Effect of approximation on the same models, considering different input size.}
    \label{fig:Effectsize}
\end{figure}

Our experimental data indicates that the robustness to write errors in state persistence operations of \acp{NN}' intermediate values is affected by input size.
\figref{fig:Effectsize} provides experimental evidence.
The results reveal that, when considering the same \ac{NN} for image classification, cases with larger input sizes exhibit milder decreases in output quality compared to those with smaller inputs.

Consider the \texttt{MobileNetV1\_0.25} benchmark as an example, using input images of dimensions 224x224 and 96x96. 
In the case of a \texttt{Q1} setting, thus with  a minimal number of bit flips, the accuracy decreases are relatively similar across all input sizes. 
As more errors occur with \texttt{Q2} and \texttt{Q3} settings, the degradation of \ac{QoR} becomes more pronounced in the network with smaller inputs. 
When errors become excessive, as in the case of a \texttt{Q4} configuration, all networks tend to fail, essentially leading to random inference. 

\fakepar{Take-away} \emph{Input size must be taken into account} alongside the configuration that determines the trade-off between energy savings and degradation of output quality, discussed in \secref{sec:resultsTradeoff}.
Smaller input sizes appear to be more vulnerable to errors in state persistence operations \emph{precisely in the parameter range that is likely most significant}.

\section{Conclusion}
\label{sec:end}

We explored the trade-off between the energy gains attainable by piloting the write current of \ac{STT-MRAM} and the reduction of output quality due to the resulting write errors.
We target battery-less \ac{IOT} devices powered by energy harvesting, where computations become intermittent because of energy failures.
We use \ac{STT-MRAM} to persist the state necessary to cross energy failures. 
We measured the energy consumption at different current settings and the output quality obtained from the execution of nine diverse benchmarks on top of seven different platforms.
Based on $1.972 \cdot 10^{12}$ experimental data points, we drew \emph{four key observations} and \emph{corresponding take-aways}: 
\begin{inparaenum}[i)]
\item the trade-off between energy saving and reduction of output quality is \emph{program-specific}; 
\item the same trade-off is a function of a \emph{platform's specific compute efficiency and power figures}; and 
\item \emph{data encoding} and \emph{input size} impact a program's resilience to errors.
\end{inparaenum}


\newpage
\bibliographystyle{ACM-Reference-Format}
\bibliography{biblio}

\end{document}